\documentclass[conference]{IEEEtran}
\usepackage{amsmath,amsfonts}
\usepackage{amsmath}
\usepackage{algorithmic}
\usepackage{algorithm}
\usepackage{array}
\usepackage[]{subfig}
\usepackage{textcomp}
\usepackage{stfloats}
\usepackage{xspace}
\usepackage{url}
\usepackage{verbatim}
\usepackage{graphicx}
\usepackage{amsthm}
\usepackage{cite}
\usepackage{cite}
\usepackage{color,soul}
\usepackage{cite}
\usepackage{color,soul}
\usepackage{multirow}
\usepackage{graphicx}
\usepackage{enumerate}
\usepackage{enumitem}
\usepackage{amsmath}
\usepackage{amsthm}
\usepackage{multirow}
\usepackage{array}
\usepackage{booktabs} 
\usepackage{bm}
\usepackage{amssymb}
\usepackage{hyperref}
\usepackage{balance}
\usepackage{xfrac}
\usepackage{mathrsfs}
\newtheorem{lemma}{\bf{Lemma}}

\newtheorem{theorem}{\bf{Theorem}}
\newtheorem{assumption}{Assumption}

\newtheorem{remark}{Remark}
\newcommand{\MATLAB}{\textsc{Matlab}\xspace}

\begin{document}
\title{Communication Reduction for Power Systems: An Observer-Based Event-Triggered Approach}

\author{
\IEEEauthorblockN{\textbf{Gabriel E. Mejia-Ruiz}\IEEEauthorrefmark{1}, 
\textbf{Yazdan Batmani}\IEEEauthorrefmark{2}, \textbf{Subhash Lakshminarayana}\IEEEauthorrefmark{3}, \textbf{Shehab Ahmed}\IEEEauthorrefmark{1},\\ \textbf{Charalambos Konstantinou}\IEEEauthorrefmark{1}}

\IEEEauthorblockA{\IEEEauthorrefmark{1}CEMSE Division, King Abdullah University of Science and Technology (KAUST)\\
\IEEEauthorrefmark{2}Department of Electrical Engineering, University of Kurdistan, Sanandaj, Iran\\
\IEEEauthorrefmark{3}School of Engineering, University of Warwick, UK
}


}

\IEEEaftertitletext{\vspace{-1\baselineskip}}
\maketitle
\begin{abstract}
The management of distributed and heterogeneous modern power networks necessitates the deployment of communication links, often characterized by limited bandwidth. This paper presents an event detection mechanism that significantly reduces the volume of data transmission to perform necessary control actions, using a scalable scheme that enhances the stability and reliability of power grids. The approach relies on implementing a linear quadratic regulator and the execution of a pair of Luenberger observers. The linear quadratic regulator minimizes the amount of energy required to achieve the control actions. Meanwhile, the Luenberger observers estimate the unmeasured states from the sensed states, providing the necessary information to trigger the event detection mechanism. The effectiveness  of the method is tested via time-domain simulations on the IEEE 13-node test feeder interfaced with inverter-based distributed generation systems and the proposed observed-based event-triggered controller. The results demonstrate that the presented control scheme guarantees the bounding of the system states to a pre-specified limit while reducing the number of data packet transmissions by $39.8$\%.
\end{abstract}

\section{Introduction}\label{Sec1}
Modern power networks include distributed generation, advanced communication, energy management technology, and intelligent infrastructure. These recent advancements in electric power grids contribute towards higher reliability, efficiency, sustainability, among other benefits \cite{zografopoulos2021cyber}. Several controllable devices, geographically separated, can be available in the power network, such as distributed generators, controllable loads, and protection equipment. A critical challenge is the proper management of the bandwidth constraints of the communication systems used for the coordinated control of all devices deployed in modern power grids \cite{G6, G8}.


Traditional control methods in power systems rely on dedicated communication channels to transmit measurement and control signals. However, modern power systems are evolving towards a more decentralized approach to control services, which often involves the use of {wireless open communication networks that 
allow different devices, systems, or applications to communicate with each other in an unrestricted way \cite{G8, 10066380, mo2018real}}. While this approach can offer benefits such as cost-effectiveness and flexibility, it also presents new challenges due to the limited bandwidth of the network. These challenges can include delays in transmission, data packet losses, and uncertainty about the parameters being measured. In such situations, traditional robust control schemes may not be effective in ensuring stable and reliable performance of the power system \cite{G5, G7}.

Recent literature reveals that there are two primary methods for reducing the number of signal transmissions in control schemes for power grids.  One approach is to maximize the sampling interval based on a time-triggered (TT) sampled data control scheme \cite{G7}. This is based on extending the sampling interval to an acceptable maximum, which does not affect the dynamic performance of the closed-loop system. Another way is to change the data transmission triggering mechanism, introducing an event-triggered (ET) communication scheme to reduce the transition of the sampled data, sending it only when a preset threshold is exceeded \cite{zhang2021fuzzy, malisoff2022event, ghodrat2021separation, batmani2019event}.

In recent years, various applications have been developed with the aim of improving the communication capabilities of control networks.  In \cite{G9}, a sampling interval of longer duration is designed for isolated hybrid power systems. A maximum admissible sampling interval for multi-area power systems is established in \cite{G10}. In contrast, a communication scheme based on sending periodically sampled data only when a preset threshold is exceeded is presented in \cite{G11}, reducing the communication overhead in multi-area power systems. In a subsequent improvement, a control architecture based on supplementary adaptive dynamic programming theory for frequency control in multi-area power systems is proposed in \cite{G12}. In \cite{song2018event}, an event-triggered observer applicable to a linear system is presented, however, {
the Zeno-free behavior
\footnote[1]{Zeno behavior describes the phenomenon in which an infinite number of transmissions occur in a finite time interval.  Zeno behavior happens when the set threshold functions in the trigger conditions are equal to zero. This phenomenon is extremely undesirable in event-triggered control which aims to save communication resources \cite{G13}.}} of the method is not analyzed. 

This paper introduces an approach that significantly reduces the data transmission in power systems. This approach is based on an event detection mechanism that triggers data transmission only when the designer-defined limits are over-passed, thus reducing the utilization of communication resources. This control approach implements a linear quadratic controller in combination with  Luenberger observers. The linear quadratic regulator optimizes the power consumption required to achieve the control commands and minimizes the error between the measured system states and those desired by the operator. Meanwhile, the Luenberger observers estimate the unmeasured states, providing full-order information to the controller. 
For observable and controllable systems, the proposed control scheme maintains the separation principle of classical control while allowing for the independent design of controllers and observers. The proposed framework can significantly reduce communication data flow while achieving almost identical control performance to that of continuous-data communication schemes. It also guarantees asymptotic stability by keeping the closed-loop system states bounded. In addition, the existence of a positive lower bound on the minimum time between events is guaranteed, avoiding Zeno behavior. 

The remaining sections of the paper are structured as follows. The theoretical background is described in Section \ref{Sec2}. Then, a detailed description of the proposed observer-based event-triggered approach is presented in \ref{sec3}. In Section \ref{Sec4}, the performance of a closed-loop system is evaluated through several scenarios, which include the simulation of the IEEE 13-node test feeder integrated with distributed generation systems based on inverters, and the implementation of the event-triggered controller using observer-based methods. Finally, concluding remarks are pointed out in Section \ref{con}.

\emph{Notations:} $\mathbb{N}=\{1,2,...\}$ represents the set of non-negative integers. $\mathbb{R}, \mathbb{R}^n$, and $\mathbb{R}^{n\times m}$ are the sets of real numbers, $n$-dimensional real vectors, and $n\times m$ real matrices, correspondingly. $I_n$ is $n\times n$ identity matrix. $0_{n\times m}$ is $n\times m$ zero matrix.  $\Vert .\Vert$ denotes a general norm. $\lambda_\mathrm{min}\{P\}$ and $\lambda_\mathrm{max}\{P\}$ are respectively the largest and smallest eigenvalues of a square matrix $P$.


\section{Problem Statement and Preliminaries}\label{Sec2}
{
We start with a generic linear time-invariant (LTI) model which can be used to represent the dynamics of several control loops in an 
electrical system around an operating point  \cite{van2016perspectives}: 
\begin{equation}
\label{ss}
\scalebox{1}{$
\begin{split}
{x}(t) = {A}{x}(t) + {B}{u}(t) \\
{y}(t)= {C}{x}(t) + {D}{u}(t),\\
\end{split}
$}
\end{equation}where $x(t) \in \mathbb{R}^{n}$, {
${u}(t) \in \mathbb{R}^{n}$ and ${y}(t) \in \mathbb{R}^{n}$ stand for the state vector, control input, and observation vectors of the system, respectively. The matrices of the linear state-space model are established by ${A}$, ${B}$, ${C}$ and ${D} \in \mathbb{R}^{n\times m}$. 
We demonstrate the effectiveness of the proposed algorithm on voltage control in distribution networks, which can be modeled as an LTI system. To derive the model, we use the popularly adopted eigensystem realization algorithm (ERA).



\subsection{Power System Identification using the Eigensystem Realization Algorithm}

The ERA is a black-box system modeling method selected for its high accuracy and computational efficiency \cite{G2}. The method to obtain the ERA solution can be summarized in the following steps:

\begin{enumerate}
\item {\it  Collect input and output data.} Exponential Chirp functions are applied as a known input sequence ($u$), as follows: 
\vspace{-3mm}
\begin{equation}
\vspace{-2mm}
\label{chirp}
u(k) = \alpha  \sin \left( \frac{2\pi f_s(r_f^k - 1)}{ \ln (r_f)} \right)
\end{equation}
where the duration of the signal sequence is $T$, the amplitude is represented by $\alpha$, $r_f$ is defined as $\left(\sfrac{f_e}{f_s}\right)^{1/T}$, and lower and upper bounds of the frequency band are $f_s$ and $f_e$, respectively. In this step, the outputs are sampled at the same rate as the control input. 

\item {\it Compute the linear relationship.} The Fourier transform applied to each input/output pair provides the equivalent impulse response from the Chirps signals. Thus, the inverse Fourier transforms of $\mathscr{F}\left(u(k)\right)$ and $\mathscr{F}\left(y(k)\right)$ results in $U(\omega)$ and $Y(\omega)$, as follows:  

\vspace{-2mm}
\begin{equation}
\label{fftinv}
\mathbf{y}(k)=\mathscr{F}^{-1}\left(\frac{Y(\omega)}{U(\omega)}\right)
\end{equation}

where $k$ is the sample indicator.


\item {\it Assemble a block-Hankel matrix.} {The data of the measured outputs are arranged in the Hankel matrix, given by:} 

\begin{equation}
    \label{mhankel}
    \scalebox{0.85}{$
    {H}_{(k-1)} = \begin{bmatrix}
    {y}_{k} & {y}_{k+1} & \cdots & {y}_{k+N} \\
    {y}_{k+1} & {y}_{k+2} & \cdots & {y}_{k+N+1} \\
    \vdots & \vdots & \ddots & \vdots \\ 
    {y}_{k+N} & {y}_{k+N+1} & \cdots & {y}_{k+2N} \\
    \end{bmatrix}
    $}
    \vspace{0.1cm}
\end{equation}where ${y}_{k} = {C}{A}^{k-1}{B}$ is the impulse response of the system.

\item {\it Perform the singular value decomposition of ${H}_0$.} The singular value decomposition of the Hankel matrix ${H}_{(k-1)}$ associated with the impulse response of the system can be obtained by:
\vspace{-2mm}
\begin{equation}
\label{eq_svd1}
    {H}_0 = {U} {\varSigma}{V}^T 
\end{equation}
where $\varSigma$ is the diagonal matrix containing the singular values of ${H}_0$. ${U}$ and ${V}$ are the left and right singular vectors, respectively.

\item {\it Find the right order of the system.} 
The system order is determined by selecting the first singular values that represent 99\% of the total energy of the system. Then, ${U}$, ${V}$ and $\varSigma$ are truncated to derive a condensed representation of the power system with reduced order.

\item {\it Compute the system matrices.} The identified model matrices of the power system under study are given by: 
\begin{equation}
	\label{eq_Aera}
	\begin{split}
	    {A} &= {\Sigma}^{-\frac{1}{2}}{U}{H}_1 {V}^T{\Sigma}^{-\frac{1}{2}}; \textup{  } {B}  ={\Sigma}^{\frac{1}{2}}{Q}^T \\
{C}  &={P}{\Sigma}^{\frac{1}{2}}; \textup{ }{D} = y(0)
\end{split}
\end{equation}
where ${H}_1$ is the shifted block-Hankel matrix of ${H}_0$. 

\end{enumerate}
}

\begin{assumption}\label{assum1}
$(A,B)$ is state controllable.    
\end{assumption}

Fig.~\ref{fig1} represents the overall diagram of the proposed control framework. Sensor data is transmitted to the controller using a communication network. In a similar way, the control action data flows from the controller to the actuator. For simplicity, in this work, we do not present a detailed model of the communication network other than accounting for the number of packet transmissions. Analysis with an in-depth communication network model will be taken up in our future work. 

The goal of this work lies in reducing communication network use. In order to achieve this objective, two strategies are exploited: (\textit{i}) The proposed approach only transmits the system output $y(t)=Cx(t)$ to the controller where $C\in\mathbb{R}^{q\times n}$ and $q<n$ decreases the number of the required packets, instead of transmitting the full vector of states. 
In this way, it is possible to reduce the required communication bandwidth and the {required energy for the transmission}. (\textit{ii}) The application of the proposed observer-based event-triggered mechanism reduces the sampling times, decreasing the use of the communication channel.


\subsection{Linear Quadratic Regulators (LQRs)}\label{subsec1}
For the considered system \eqref{ss}, the challenge is to construct a control law that minimizes:  
\begin{eqnarray}\label{lqr}
J = \frac{1}{2}\int_0^\infty \! \big(x^\mathrm{T}(t)Qx(t)+u^\mathrm{T}(t)Ru(t)\big) \, \mathrm{d}t 
\end{eqnarray}
where $Q\geq 0$ and $R>0$ are user-defined matrices with appropriate dimensions. The well-known solution to this optimal control problem is found using the linear quadratic regulator (LQR) approach \cite{G1, G2}. The state-feedback control to minimize $J$ \eqref{lqr} is: 
\vspace{-2mm}
\begin{equation}
u(t)=-Kx(t),
\end{equation}
where $K\in\mathbb{R}^{m\times n}$ is given by $K=R^{-1}B^\mathrm{T}P$,  and $P$ is the unique symmetric positive-definite solution of the Riccati equation: 
\begin{eqnarray}\label{riclqr}
\label{e6}
  \begin{split}
A^\mathrm{T}P+PA-PBR^{-1}B^\mathrm{T}P+Q=0_{n\times n}.
\end{split}
\end{eqnarray}

\begin{lemma}\label{lemma1}
{If $(A,B,Q^{1/2})$ is stabilizable and detectable, then the real parts of all the eigenvalues of $A-BK$ are strictly negative, i.e., $A-BK$ is a Hurwitz matrix \cite{G2}.}  
\end{lemma}

\subsection{Linear Full-Order Observers}\label{subsec2}
Consider the linear system \eqref{ss}. Let $\hat{x}(t) \in \mathbb{R}^{n}$ be the estimated state of the system. A classical choice to estimate $x(t)$ from $y(t)$ is by following the Luenberger observer \cite{luenberger1964observing}:
\begin{eqnarray}\label{e4}
  \begin{split}
\dot{\hat{x}}(t)&=A\hat x(t)+Bu(t)+L(y(t)-C\hat{x}(t)),
\end{split}
\end{eqnarray}
where $L \in \mathbb{R}^{n\times q}$ is the observer gain. If every eigenvalue of $A-LC$ has a strictly negative part, the estimation error $e(t)=x(t)-\hat{x}(t)$ tends to zero when $t\rightarrow 0$. 

\begin{figure}[t]
\centering
\includegraphics[scale=0.8]
{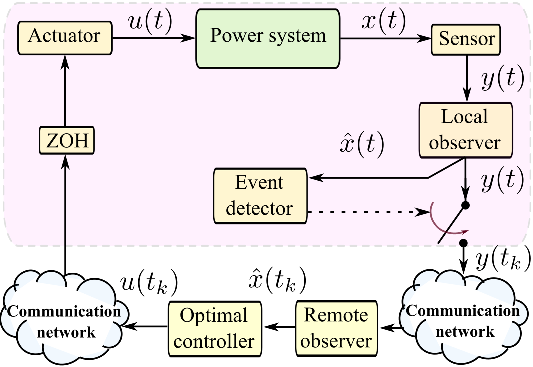}\setlength{\belowcaptionskip}{-8pt}
 \caption{Overall diagram of event-triggered observer-based control.}\label{fig1}
 \vspace{-3mm}
 \end{figure}

\begin{lemma}\label{lemma2}
{$L$ can be found to assign all the eigenvalues of $A-LC$ to arbitrarily locations if and only if the system \eqref{ss} is state observable \cite{chen1998linear}.}    
\end{lemma}

To find $L$, the solution $S=S^\mathrm{T}>0_{n\times n}$ of the Riccati equation:
\begin{eqnarray}\label{ricc}
\label{e6}
  \begin{split}
AS+SA^\mathrm{T}-SC^\mathrm{T}V^{-1}CS=-W,
\end{split}
\end{eqnarray}
can be obtained where $V>0_{q\times q}$ and $W\geq 0_{n \times n}$ are determined by the designer. It is possible to prove that if $(A,C)$ is  observable and $L=SC^\mathrm{T}V^{-1}$, $A-LC$ is a Hurwitz matrix \cite{simon2006optimal}.

\section{Proposed Method}\label{sec3}
As depicted in Fig. \ref{fig1}, an observer is established after the first point in the communication network, where the data is transmitted from the system to the {controller}. The observer aims to estimate $x(t)$ using $y(t_\mathrm{k})$. We will refer to this observer as a ``remote observer''. Here, $t_\mathrm{k}$, $\mathrm{k}\in \mathbb{N}$ stands for the instants when $y(t)$ is sent to the observer. The local observer, which is a copy of the remote observer and is installed near the system, is utilized to reproduce $\hat{x}(t)$. 


Since $y(t)$ is transmitted to the observers (local and remote) only at $t_\mathrm{k}$, $k \in \mathbb{N}$, \eqref{e4} can be rewritten as follows for $t \in [t_\mathrm{k},t_\mathrm{k+1})$:
\begin{eqnarray}\label{e5}
  \begin{split}
\dot{\hat{x}}(t)&=A\hat x(t)+Bu(t_\mathrm{k})+LC(x(t_\mathrm{k})-\hat x(t)).
\end{split}
\end{eqnarray}

\noindent Moreover, during $t \in [t_\mathrm{k},t_\mathrm{k+1})$, $u(t)=-K\hat{x}(t_\mathrm{k})$ and hence, the system defined \eqref{ss} evolves as: 
\begin{eqnarray}\label{ee5}
  \begin{split}
\dot{x}(t)&=Ax(t)-BK\hat{x}(t_\mathrm{k})\\
&=(A-BK)x(t)+BK(x(t)-\hat{x}(t_\mathrm{k}))
\end{split}
\end{eqnarray}

\noindent Adding and subtracting $\hat{x}(t)$ to the last part of the right hand side of \eqref{ee5} leads to:
\begin{align}\label{eee3}
\dot{x}(t)=(A-BK)x(t)+BKe(t)+BK(\hat{x}(t)-\hat{x}(t_\mathrm{k})).
\end{align}

\noindent Using \eqref{e5} and \eqref{ee5}, we have:
\begin{eqnarray}\label{ee4}
\dot{e}(t)=(A-LC)e(t)+L(y(t)-y(t_\mathrm{k})).
\end{eqnarray}

Let us define $X(t)=\begin{bmatrix}
x^\mathrm{T}(t)& e^\mathrm{T}(t)
\end{bmatrix}^\mathrm{T}\in \mathbb{R}^{2n}$. Using \eqref{eee3} and \eqref{ee4}, the dynamic of $X(t)$ is described as: 
\begin{equation}\label{EEa}
\dot{X}(t)=\tilde{A}X(t)+\psi(t,t_\mathrm{k}),
\end{equation}
where 
\begin{eqnarray}\label{hh}
\tilde{A}=\begin{bmatrix}
A-BK & BK\\ 0_{n\times n} & A-LC
\end{bmatrix}\in \mathbb{R}^{2n\times 2n},
\end{eqnarray}
and 
\begin{eqnarray}\label{psi}
\psi(t,t_\mathrm{k})=\begin{bmatrix}
BK(\hat{x}(t)-\hat{x}(t_\mathrm{k}))\\L(y(t)-y(t_\mathrm{k}))
\end{bmatrix}\in \mathbb{R}^{2n}.
\end{eqnarray}

\noindent The problem is to find $t_\mathrm{k}$, $\mathrm{k}\in \mathbb{N}$ such that \eqref{EEa} is asymptotically stable, \textit{i.e.}, $x(t), e(t)\rightarrow 0$ when $t\rightarrow \infty$. 

Following Lemmas \ref{lemma1} and \ref{lemma2} and under Assumption \ref{assum1}, $A-BK$ and $A-LC$ are Hurwitz for $K=R^{-1}B^\mathrm{T}P$ and $L=SC^\mathrm{T}V^{-1}$. On the other hand, since $\tilde{A}$ is a lower triangular matrix, it is also Hurwitz. A unique solution $\tilde{P}=\tilde{P}^\mathrm{T}>0$ of the following Lyapunov equation exists for any $\tilde{Q}=\tilde{Q}^\mathrm{T}>0$:  
\begin{equation}\label{lya}
\tilde{A}^\mathrm{T}\tilde{P}+\tilde{P}\tilde{A}+\tilde{Q}=0_{2n\times 2n}.
\end{equation}

Taking into account the Lyapunov candidate function $V(X(t))=X^\mathrm{T}(t)\tilde{P}X(t)$, its derivative is:
\begin{eqnarray}
\label{e9}
  \begin{split}
\dot{V}(X(t))&=\dot{X}^\mathrm{T}(t)\tilde{P}X(t)+X^\mathrm{T}(t)\tilde{P}\dot{X}(t).
\end{split}
\end{eqnarray}

\noindent Using \eqref{EEa} in \eqref{e9}, we have: 
\begin{eqnarray}\label{ee9}
\begin{split}
\dot{V}(X(t))&={X}^\mathrm{T}(t)(\tilde{A}^\mathrm{T}\tilde{P}+\tilde{P}\tilde{A})X(t)\\
&+X^\mathrm{T}(t)\tilde{P}\psi(t,t_\mathrm{k})+\psi^\mathrm{T}(t,t_\mathrm{k})\tilde{P}X(t).
\end{split}
\end{eqnarray}

\noindent In a time-triggered framework, $\dot{V}(X(t))=-{X}^\mathrm{T}(t)\tilde{Q}X(t)$ since $\psi(t,t_\mathrm{k})=0$ for $t\geq 0$. In the proposed method, however, a weaker rate of decrease in $V(X(t))$ is considered by imposing the inequality: 
\begin{eqnarray}\label{eee9}
\dot{V}(X(t))\leq -\sigma X^\mathrm{T}(t)\tilde{Q}X(t),
\end{eqnarray}
where $\sigma\in (0,1]$ is a constant, called an event-triggered factor. Substituting \eqref{ee9} in \eqref{eee9}, we have: 
\begin{eqnarray}\label{ee12}
(\sigma-1){X}^\mathrm{T}(t)\tilde{Q}X(t)+2{X}^\mathrm{T}(t)\tilde{P}\psi(t,t_\mathrm{k})\leq 0,
\end{eqnarray}
that is rearranged as: 

\begin{eqnarray}\label{e7}
  \begin{split}
&\begin{bmatrix}
X^\mathrm{T}(t) & \psi^\mathrm{T}(t,t_\mathrm{k})
\end{bmatrix}\Phi\begin{bmatrix}
X(t) \\ \psi(t,t_\mathrm{k})
\end{bmatrix}\leq 0,
\end{split}
\end{eqnarray}
where $\Phi\in \mathbb{R}^{4n\times 4n}$ is: 
\begin{equation}\label{Psi}
\Phi=\begin{bmatrix}
(\sigma-1)\tilde{Q} & \tilde{P} \\ \tilde{P} & 0_{2n\times 2n}
\end{bmatrix}.
\end{equation}

\begin{theorem}[]\label{theorem1}
Assume \eqref{ss} is state controllable and $u(t)=-K\hat{x}(t_\mathrm{k})$ is applied where  $K=R^{-1}B^\mathrm{T}P$. Assume further that $L=SC^\mathrm{T}V^{-1}$ is selected where $S$ is the solution of \eqref{ricc}. If the measured output $y(t)$ is transmitted using the network when \eqref{e7} is violated, then $x(t)$ and $e(t)$ converge to zero asymptotically. 
\begin{proof}
According to \eqref{eee9} and due to \eqref{e7}, $\dot{V}(X(t))$ is always negative. Therefore,  $X(t)=\begin{bmatrix} 
x^\mathrm{T}(t) & e^\mathrm{T}(t)
\end{bmatrix}^\mathrm{T}$ is asymptotically stable.     
\end{proof}
\end{theorem}

\begin{remark}\label{remark1}
From the structure of $\tilde{A}$ in \eqref{hh}, it can be concluded that the separation principle of classical control holds.    
\end{remark}

Paying attention to \eqref{e7}, the triggering times are: 
\begin{align}\label{eew}
t_\mathrm{k+1}=\mathrm{inf}\{t>t_\mathrm{k} \:\big|\:
[X^\mathrm{T}(t)\quad \psi^\mathrm{T}(t,t_\mathrm{k})]\Phi\begin{bmatrix}
X(t) \\ \psi(t,t_\mathrm{k})
\end{bmatrix}\geq 0\}
\end{align}
where $\mathrm{k}\in \mathbb{N}$, $t_1=0$, and $\psi(t,t_\mathrm{k})$ and $\Phi$ are defined in \eqref{psi} and \eqref{Psi}, respectively. To considerably reduce transmissions over the network while maintaining the $X(t)$ bound, the event-detector mechanism \eqref{eew} can be systematically equipped with additional conditions. In other words, there is no need to utilize the network if the state $X(t)$ is close enough to zero. However, this issue may lead to some errors in the estimation and control. For instance, the condition $\Vert X(t)\Vert> \epsilon$ could be added to the event-detector block with $\epsilon>0$ as a constant. In this situation, \eqref{eew} is replaced with: 
\begin{align}\label{eew2}
t_\mathrm{k+1}=\mathrm{inf}\{t>t_\mathrm{k} &\:\big|\:
\begin{bmatrix}
X^\mathrm{T}(t) & \psi^\mathrm{T}(t,t_\mathrm{k})
\end{bmatrix}\Phi\begin{bmatrix}
X(t) \\ \psi(t,t_\mathrm{k})
\end{bmatrix}\geq 0\nonumber\\
 &\&\, \quad \Vert X(t)\Vert >\epsilon\}.
\end{align}

If the added condition causes instability of the system, $\Vert X(t)\Vert$ increases. Therefore, this condition will be relaxed and the triggering conditions \eqref{eew2} and \eqref{eew} will be the same. In this situation, according to Theorem \ref{theorem1}, $\Vert X(t)\Vert$ will be decreased. Hence, adding $\Vert X(t)\Vert> \epsilon$ to the event-triggered mechanism may cause some estimation and control errors. However, the adverse effects of these errors on the system performance can be reduced by choosing appropriate values for $\epsilon$. In order to select a proper value for this parameter which leads to a desirable compromise between the performance and usage of the network, the designer can start with an arbitrarily small value of $\epsilon$. If the performance of the system is satisfactory, larger values of $\epsilon$ can be also considered. If not, the designer should test smaller values of these parameters until a good compromise is achieved between communication usage and performance.



\begin{theorem}\label{theorem2}
The system state and estimation error are globally uniformly and ultimately bounded.
\begin{proof}
For $V(X(t))=X^\mathrm{T}(t)\tilde{P}X(t)$, we have:
\begin{align*}
\alpha_1(\Vert X(t)\Vert)&\triangleq\lambda_\mathrm{min}\{\tilde{P}\}\Vert X(t)\Vert^2\leq V(X(t))\leq \alpha_2(\Vert X(t)\Vert)\nonumber\\
& \triangleq\lambda_\mathrm{max}\{\tilde{P}\}\Vert X(t)\Vert^2.
\end{align*}
Besides, using \eqref{eee9}, one can conclude that $\dot{V}(X(t))\leq -W_3(X(t))$ for $\Vert X(t)\Vert \geq \mu=\epsilon^+$ where $W_3(X(t))=\sigma X^\mathrm{T}(t)\tilde{Q}X(t)$. As $\alpha_1(\Vert X(t)\Vert)$ is a class $\mathcal{K}_\infty$ function and from Theorem 4.18 in \cite{khalil2002nonlinear}, with the ultimate bound $\alpha_1^{-1}(\alpha_2(\mu))=\sqrt{\lambda_\mathrm{max}\{\tilde{P}\}/\lambda_\mathrm{min}\{\tilde{P}\}}\mu$, $X(t)$ is globally uniformly bounded.         
    \end{proof}
\end{theorem}
 
\begin{theorem}\label{theorem3}
There exists $\tau>0$ such that $t_\mathrm{k+1}-t_\mathrm{k}\geq \tau$ for $\mathrm{k}\in \mathbb{N}$.  
\begin{proof}
Paying attention to the event-triggering condition \eqref{eew2}, let us consider two cases:\\ (1) $\Vert X(t)\Vert<\epsilon$ for $t=t_\mathrm{k}^+$: In this situation, a limited or even unlimited amount of time is needed to reach $\Vert X(t)\Vert =\epsilon$. If $\Vert X(t)\Vert <\epsilon$ for $t>t_\mathrm{k}$, no new event is triggered.  (see case study in  Section \ref{Sec4}). \\
(2) $\Vert X(t)\Vert>\epsilon$ for $t=t_\mathrm{k}^+$: The rest of the proof is dedicated to this case. 

Define $\theta(t)=\Vert \psi(t,t_\mathrm{k})\Vert/\Vert X(t)\Vert$, $X(t)\neq 0$. The following dynamic for $\theta(t,t_\mathrm{k})$ is obtained:
\begin{align}
&\frac{\mathrm{d}\theta(t)}{\mathrm{d}t}=\frac{\mathrm{d}}{\mathrm{d}t}\bigg(\frac{(\psi^\mathrm{T}(t,t_\mathrm{k})\psi(t,t_\mathrm{k}))^{\frac{1}{2}}}{(X^\mathrm{T}(t)X(t))^{\frac{1}{2}}}\bigg)\label{eein}\\
&=\frac{(\psi^\mathrm{T}(t,t_\mathrm{k})\psi(t,t_\mathrm{k}))^{-\frac{1}{2}}\psi^\mathrm{T}(t,t_\mathrm{k})\dot{\psi}(t,t_\mathrm{k})(X^\mathrm{T}(t)X(t))^{\frac{1}{2}}}{X^\mathrm{T}(t)X(t)}\nonumber\\
&-\frac{(X^\mathrm{T}(t)X(t))^{-\frac{1}{2}}X^\mathrm{T}(t)\dot{X}(t)(\psi^\mathrm{T}(t,t_\mathrm{k})\psi(t,t_\mathrm{k}))^{\frac{1}{2}}}{X^\mathrm{T}(t)X(t)}\nonumber\\
&=-\frac{\psi^\mathrm{T}(t,t_\mathrm{k})\dot{\psi}(t,t_\mathrm{k})}{\Vert \psi(t,t_\mathrm{k})\Vert \Vert X(t)\Vert}-\frac{X^\mathrm{T}(t)\dot{X}(t)}{\Vert X(t)\Vert \Vert X(t)\Vert}\frac{\Vert \psi(t,t_\mathrm{k})\Vert}{\Vert X(t)\Vert}\nonumber
\end{align}

The following inequality is achieved based on \eqref{eein} and using the triangle inequality:
\begin{eqnarray}\label{eein1}
\begin{split}
\frac{\mathrm{d}\theta(t)}{\mathrm{d}t}&\leq \frac{\Vert\dot{\psi}(t,t_\mathrm{k})\Vert}{\Vert X(t)\Vert}+\frac{\Vert\dot{X}(t)\Vert}{\Vert X(t)\Vert}\frac{\Vert \psi(t,t_\mathrm{k})\Vert}{\Vert X(t)\Vert},
\end{split}
\end{eqnarray}

Due to Theorem \ref{theorem1}, $X(t)$ is bounded and as a result, there is a positive real number $\beta$ such that $\Vert \dot{\psi}(t,t_\mathrm{k})\Vert\leq \beta$. Since $\Vert X(t)\Vert>\epsilon$, there exists $\gamma>\beta/\epsilon$ such that: 
\begin{eqnarray}\label{eein13}
\begin{split}
\frac{\Vert\dot{\psi}(t,t_\mathrm{k})\Vert}{\Vert X(t)\Vert}\leq \gamma.
\end{split}
\end{eqnarray}
Using \eqref{EEa} and the triangle inequality, we have:
\begin{align}\label{eer}
\Vert \dot{X}(t)\Vert & \leq \Vert \tilde{A}\Vert \Vert X(t)\Vert+\Vert \psi(t,t_\mathrm{k})\Vert \leq (1+\Vert \tilde{A}\Vert)\big(\Vert X(t)\Vert\nonumber\\
&+\Vert \psi(t,t_\mathrm{k})\Vert\big).
\end{align}
Using inequalities \eqref{eein1}, \eqref{eein13}, and \eqref{eer} leads to: 
\begin{eqnarray}\label{eint}
\begin{split}
\frac{\mathrm{d}\theta(t)}{\mathrm{d}t}\leq \gamma+(1+\Vert \tilde{A}\Vert)\big( 1+\theta(t)\big)\theta(t).
\end{split}
\end{eqnarray}
Since $\theta(t)$ is not negative, the following inequality is obtained by adding $\alpha(1+\theta(t))$ to the right-hand side of \eqref{eint}:
\begin{eqnarray}\label{eint2}
\frac{\mathrm{d}\theta(t)}{\mathrm{d}t}\leq \gamma+\alpha(1+\theta(t))^2,
\end{eqnarray}
where $\alpha =1+\Vert \tilde{A}\Vert$. Now, let us take integral of \eqref{eint2} from $t_\mathrm{k}$ to $t_\mathrm{k+1}^-$ as:
\begin{eqnarray}\label{int}
\int_{\theta(t_\mathrm{k})=0}^{\theta(t_\mathrm{k+1}^-)} \frac{\mathrm{d}\big(\theta(t)\big)}{\gamma+\alpha(1+\theta(t))^2}\leq \int_{t_\mathrm{k}}^{t_\mathrm{k+1}^-} \mathrm{d}t.
\end{eqnarray}
From \eqref{int}, the following inequality is obtained as $\theta(t_\mathrm{k})=0$:
\begin{eqnarray}\label{iiiu}
\begin{split}
\sqrt{\alpha\gamma}(t_\mathrm{k+1}^--t_\mathrm{k})&\geq \tan^{-1}\big(\sqrt{\alpha\gamma^{-1}}(1+\theta(t_\mathrm{k+1}^-))\big)\\
&-\tan^{-1}\big(\sqrt{\alpha\gamma^{-1}}\big).
\end{split}
\end{eqnarray}
Based on \eqref{ee12}, at $t=t_\mathrm{k+1}^-$, the following equality holds:
\begin{align}\label{xxxx}
2{X}^\mathrm{T}(t_\mathrm{k+1}^-)\tilde{P}\psi(t_\mathrm{k+1}^-,t_\mathrm{k})=(1-\sigma){X}^\mathrm{T}(t_\mathrm{k+1}^-)\tilde{Q}X(t_\mathrm{k+1}^-).
\end{align}

Using \eqref{xxxx}, the \textit{Cauchy-Schwarz} inequality $|v^\mathrm{T}w|\leq \Vert v\Vert \Vert w\Vert$ \cite{mitrinovic2013classical}, and since $w^\mathrm{T}\tilde{Q} w\geq \lambda_\mathrm{min}\{\tilde{Q}\}\Vert w\Vert ^2$, the following inequality is achieved: 
\begin{eqnarray*}
\begin{split}
\Vert{X}(t_\mathrm{k+1}^-)\Vert \Vert \tilde{P}\Vert \Vert\psi(t_\mathrm{k+1}^-,t_\mathrm{k})\Vert &\geq X^\mathrm{T}(t_\mathrm{k+1}^-)\tilde{P}\psi(t_\mathrm{k+1}^-,t_\mathrm{k})\\
& \geq \frac{1-\sigma}{2}\lambda_\mathrm{min}\{\tilde{Q}\}\Vert X(t_\mathrm{k+1}^-)\Vert^2
\end{split}
\end{eqnarray*}
which leads to: 
\begin{eqnarray}\label{yy}
\theta(t_\mathrm{k+1}^-)=\frac{\Vert\psi(t_\mathrm{k+1}^-,t_\mathrm{k})\Vert}{\Vert X(t_\mathrm{k+1}^-)\Vert}\geq \frac{(1-\sigma)\lambda_\mathrm{min}\{\tilde{Q}\}}{2\Vert \tilde{P}\Vert}>0.
\end{eqnarray}
From \eqref{iiiu} and \eqref{yy}, $t_\mathrm{k+1}^--t_\mathrm{k}$ is strictly positive for $\mathrm{k}\in \mathbb{N}$ since the function $\tan(.)$ is a strictly increasing function. The proof is completed.

\end{proof}
\end{theorem}

\section{Simulation Results for an Inverter-Based Distribution System}\label{Sec4} 

\begin {figure}[!t]
\centering
\includegraphics[width=3in]{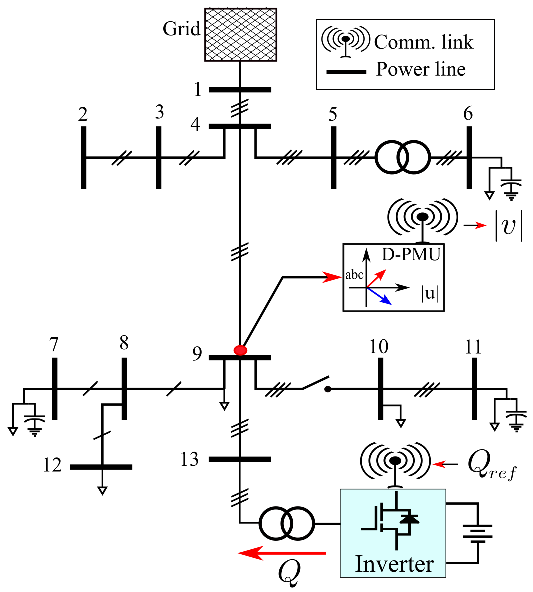}
\vspace{-1mm}
\caption{One-line diagram of the IEEE 13-node feeder interfaced with inverter-based generators.
\label{IEEE13nodes}}
\vspace{-4mm}
\end{figure}

The feasibility of the proposed event-trigger-based control structure is evaluated using the 13-node IEEE test feeder, whose single-line diagram is shown in Fig. \ref{IEEE13nodes}. {This network operates at $4.16$ kV, it is characterized by being short, relatively highly loaded, with a single voltage regulator at the substation, shunt capacitors, and unbalanced loads \cite{G2}.}
This network is equipped with a distributed generation scheme based on inverters connected to bus 13 via a step-up transformer. The inverter assumes the role of the actuator in the control scheme, injecting the optimal reactive power calculated by the LQR. Voltage amplitude measurements are provided by a D-PMU installed at node 9. 
Application of the ERA black-box system identification method yields the reduced-order model with six states shown in \eqref{A} and \eqref{BCD}. The input-output data for model construction is collected by simultaneous measurement of the {reactive power set-point on the inverter side and the voltage amplitude sensed by the D-PMU at node 9}. While running the system identification algorithm, the reactive power setpoint in the inverter-based generator deployed at node 13 is modulated with the Chirp signal.


\begin{equation}\label{A}
\resizebox{.85\hsize}{!}{$
A =\begin{bmatrix}
-41.1&-14.5&38.8&-14.6&-11.7&-6.6\\
14.5&2.7&1.8&-0.3&-0.2&-0.2&\\
-38.8&1.8&-40.5&63.6&25.4&17.4\\
-14.6&0.25&-63.6&-24.6&-58.3&-19.1\\
11.7&-0.17&25.4&58.3&-35.1&-54.9\\
-6.57&0.16&-17.4&-19.1&54.92&-45.4
\end{bmatrix} $}
\end{equation}

\begin{equation}\label{BCD}
B =\begin{bmatrix}
-0.05\\ 0.002\\ -0.02\\ -0.01\\ 0.008\\ -0.004\\ \end{bmatrix}, C' =\begin{bmatrix}
-2.69e^{-05}\\-7.5e^{-07}\\1.04e^{-05}\\-4.86e^{-06}\\-3.55e^{-06}\\-1.99e^{-06}\\
\end{bmatrix}, D=0
\end{equation}

With the inputs and outputs determined by the sensors and actuators deployed over the network, the open-loop system is asymptotically stable, controllable and observable, making the implementation of the proposed method feasible. The closed-loop system was coded in \MATLAB, based on the Euler method with a sampling frequency of $1$ kHz. In this work, $Q=W=I_6$ and $R=V=1$. 
By using these matrices selected by the designer and the application of the procedures described in \ref{subsec1} and \ref{subsec2}, the LQR controller gain and the observer gains are computed as: 
\begin{align*}
    K=[0.06, 0.26, 0.03, 0.02, -0.009, 0.008]
\end{align*}
\begin{align*}
    L=[-2.9e^{-05}, 1e^{-04}, 1.6e^{-05}, -8e^{-06}, -5e^{-06}, -4.3e^{-06}]^\mathrm{T}
\end{align*}
The initial conditions are $x(0)=[
3\quad 5\quad 5\quad 0\quad 0\quad 0]^\mathrm{T}$, $\hat{x}(0)=[2\quad 3\quad 2\quad 0\quad 0\quad 0]^\mathrm{T}$, $\sigma=0.95$, and $\epsilon = 0.1$.

\begin {figure}[t]
\centering
\includegraphics[width=3.3 in]{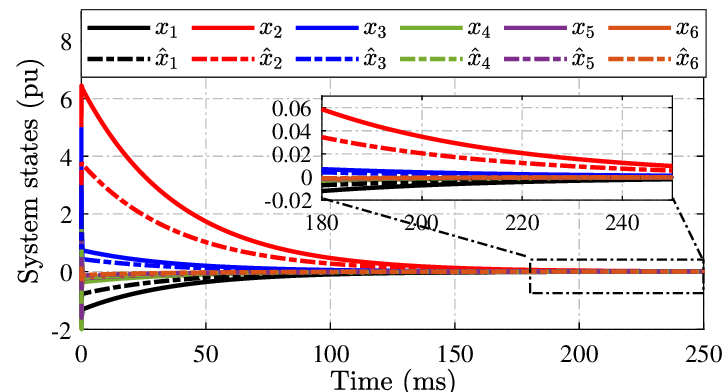}
\vspace{-1mm}
\caption{System states and their estimations.
\label{states}}
\vspace{-3mm}
\end{figure}






The system state and their estimations for the closed-loop system are depicted in Fig. \ref{states}. The presented state feedback method, based on the optimal gain $K$ and the Luenberger observer, assumes that all state variables are measurable or observable and its objective is to drive the output towards zero, rejecting disturbances. The stability and transient response characteristics of the closed-loop system are determined by the characteristic values of the $(A-BK)$ matrix. The optimal choice of matrix $K$ for the LQR controller allows the $(A-BK)$ matrix to be asymptotically stable, making it possible to drive $x(t)$ to tend to $0$ when $t$ tends to infinity. With this state feedback, the real parts of the closed-loop poles are negative, evidencing the stability of the system. 
The dynamic behavior of the observer's error vector is determined by the eigenvalues of the $(A-LC)$ matrix. For this example, the $(A-LC)$ matrix is stable and the error ($e=x-\hat{x}$) converges to zero for any initial error vector, as shown in Fig. \ref{states}.

\begin {figure}[!t]
\centering
\includegraphics[width=3.3 in]{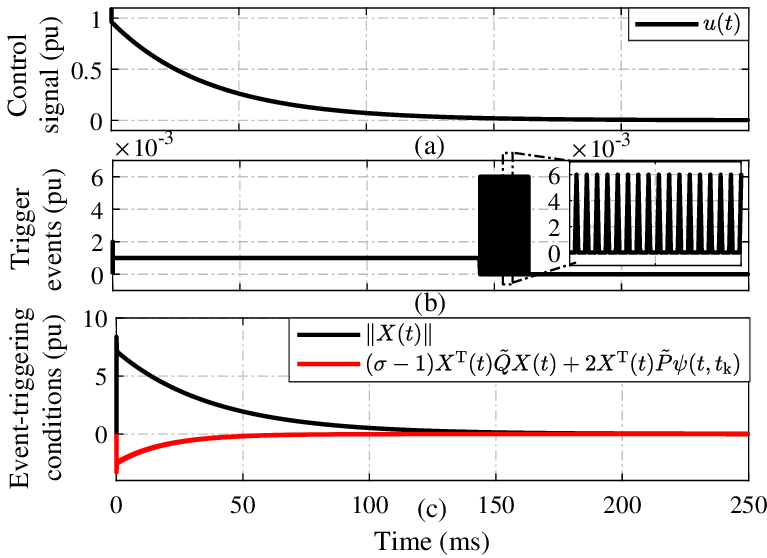}
\vspace{-1mm}
\caption{Amplitudes of the (a) control signal $u(t)$, 
(b) inter-execution times, and (c) triggered events following the times given by \eqref{eew2}.
\label{amplitudes}}
\vspace{-3mm}
\end{figure}

The computed control signal and the event instants are depicted in Fig. \ref{amplitudes}(a) and Fig. \ref{amplitudes}(b), respectively. Meanwhile, the evolution of the event-triggering conditions, computed in \eqref{eew2}, is shown in Fig. \ref{amplitudes}(c). At the beginning of the simulation, the error between the measured output with respect to the expected zero drives the control signal to 1. Subsequently, the feedback control action reduces the error asymptotically to zero within $250ms$, as shown in Fig. \ref{amplitudes}(a). When the proposed event detection system is operating, {the application of the trigger conditions reduces the number of samples by $39.8$\%, as shown in Fig. \ref{events}.} In those results, the rate of data exchange is decreasing when the system is going to the steady state, as is shown between $130ms$ and $170ms$ of the simulation in Fig. \ref{amplitudes}(b). The trigger condition $\Vert X(t)\Vert > \epsilon$ becomes inactive after $164ms$, while the second condition remains active until $187ms$, as shown in Fig. \ref{amplitudes}(c).

\begin {figure}[!t]
\centering
\includegraphics[width=3.1 in]{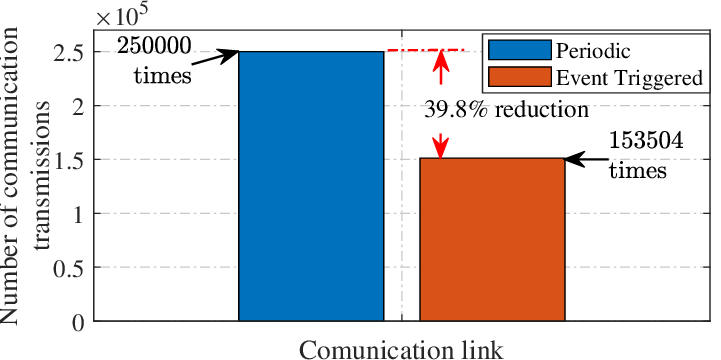}
\vspace{-1mm}
\caption{Periodic
communication transmissions and event-triggered communication comparison.
\label{events}}
\vspace{-3mm}
\end{figure}

{Fig. \ref{sigma} shows the impact of $\sigma$ on the number of communication events. The number of trigger events rises as $\sigma$ increases, following the trigger rule depicted in \eqref{ee12}. The resulting increase in the transmitted samples reduces the availability of the communication resources deployed. The suitable selection of $\sigma$ significantly impacts the usage efficiency of communication resources.} Based on the aforementioned results, the proposed controller is able to regulate the state variables of the power electric system, significantly reducing the use of communication channels. 

\begin {figure}[!t]
\centering
\includegraphics[width=3.3 in]{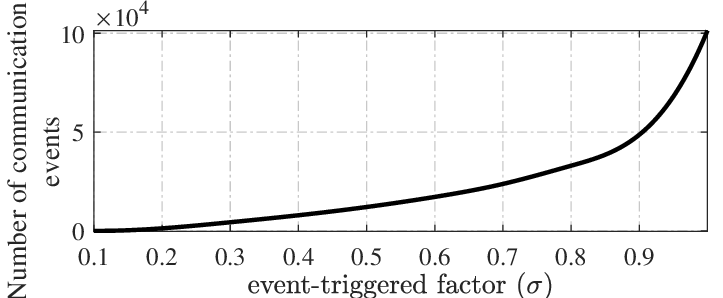}
\vspace{-1mm}
\caption{Dependence between the number of communication events and $\sigma$ selection.
\label{sigma}}
\vspace{-4mm}
\end{figure}
}

\section{Conclusions}\label{con}
The presented work combines two approaches to achieve asymptotic convergence of the variables of interest, and simultaneously reduce the utilization of communication resources between the controller and the sensors/actuators deployed over the electric power systems. An optimal state-feedback control scheme regulates the system variables, minimizing the energy required to reach the steady-state setpoints. Luenberger observers estimate the unmeasured states over the system by supplying full-order information to the controller. This framework also includes an event detection mechanism that reduces the utilization of communication resources by up to $39.8$\%. The simulations of the proposed method demonstrate that the closed-loop control framework allows for asymptotically bounding of the system states, while the trigger event handler mechanism reduces the communication bandwidth consumption without constraining the stability and excluding the Zeno behavior. This reveals the trade-off between control performance and communication resource utilization through {the} proper selection of parameters in the event trigger conditions. The proposed control scheme is highly beneficial for improving reliability, scalability, and flexibility in modern power grids. In our future work, we will explicitly incorporate the communication network model and co-design the control and communication parameters to achieve efficient event-triggered control in power systems.

\bibliographystyle{IEEEtran} 
\bibliography{Refs}
\end{document}